\documentclass[twocolumn,showpacs,preprintnumbers,amsmath,amssymb]{revtex4}
\usepackage[dvips]{graphicx}
\usepackage{dcolumn}
\usepackage{color}
\usepackage{amsmath}

\def\ket#1{  \left\vert  #1   \right\rangle   }
\def\bra#1{  \left\langle  #1   \right\vert   }
\def\etal{\textit{et al.}}
\def\cor#1{{{#1}}}

\begin{document}

\title{Symmetries of Multipartite Entanglement Evolution in Many-Sided Local Channels}

\author{Michael Siomau}
 \email{m.siomau@gmail.com}
 \affiliation{Physics Department, Jazan University, P.O.~Box 114, 45142 Jazan, Kingdom of Saudi Arabia}

\date{\today}

\begin{abstract}
Symmetries of the initial state of a quantum system and the quantum
channels, which simultaneously affect parts of the system, can
significantly simplify the description of the entanglement
evolution. Using concurrence as the entanglement measure, we study
the entanglement evolution of few qubit systems, when each of the
qubits is affected by a local channel independently of the others.
We show that, for low-rank density matrices of the final quantum
state, such complex entanglement dynamics can be completely
described by a combination of independent factors representing the
evolution of entanglement of the initial state, when just one of the
qubits is affected by a local channel.
\end{abstract}


\maketitle

\section{Introduction}

Current development of quantum technologies suggests quantum
entanglement as the exclusive resource for many potential
applications, such as quantum teleportation, superdense coding,
quantum cryptography and quantum computing \cite{Horodecki:09}.
However, apart from entanglement generation
\cite{Plenio:99,Jennewein:00,Edamatsu:04} and detection
\cite{Guehne:09}, successful practical utilization of
entanglement-based quantum technologies demands efficient protocols
for entanglement protection from detrimental environmental influence
\cite{Akulin:05} and its recovery after a possible partial loss
\cite{Sun:10, Kim:12, Siomau:12a}. The construction of the
protocols, in turn, requires exact methods for entanglement
quantitative description as well as clear understanding of the
fundamental laws of the entanglement evolution. The lack of accurate
entanglement measure for multipartite entangled systems
\cite{Horodecki:09} imposes serious limitations on our ability to
describe complex entanglement dynamics. Therefore, any realistic
situation, when the description of the entanglement evolution of a
multipartite quantum system can be simplified, is of great practical
importance.

An important example of such a simplified description of complex
entanglement dynamics was given by Konrad \etal{} \cite{Konrad:08}.
It was shown, in particular, that the entanglement evolution of an
arbitrary pure two-qubit state can be completely described by two
factors, which are given by the initial entanglement of the pure
state and the entanglement dynamics of the maximally entangled
state. Subsequently, this result has been extended to the cases of
high-dimensional bipartite systems
\cite{Tiersch:08,Li(d):09,Gour:10}, mixed initial states
\cite{Yu:08} and multiqubit systems \cite{Siomau:10b}. All these
results, however, were obtained under assumption that just one
subsystem of the entangled system undergoes the action of an
environmental channel (i.e. the system is affected by a single-sided
channel). In practice, however, it is often required to distribute
parts of an entangled system between several remote recipients
\cite{Kimble:08}. In this case, each subsystem is coupled locally
with some environmental channel, i.e. the quantum system is the
subject of many-sided channels. Recently, we analyzed the
entanglement dynamics of initially pure three-qubit
Greenberger-Horne-Zeilinger (GHZ) state, when each qubit is
simultaneously affected by a noisy channel \cite{Siomau:12b}. We
showed that, in some cases, the entanglement dynamics of the
three-qubit system in many-sided channels can be completely
described by factors, which represent the evolution of the entangled
system in single-sided channels. Similar result has been
independently obtained by Man \etal{} \cite{Man:12} for generalized
multiqubit GHZ states. Moreover, using so-called G-concurrence
\cite{Gour:05} as the entanglement measure, Gheorghiu and Gour
\cite{Gheorghiu:12} have recently shown that the average loss of
entanglement induced by many-sided local channels is independent on
the initial state and is completely defined by the local channels.

In this paper, we analyze the entanglement evolution of two-, three-
and four-qubit systems affected by local many-sided channels. Using
Wootter's concurrence \cite{Wootters:98} for two qubits and its
extension to higher dimensions \cite{Ou:08} for multiqubit systems,
we show that, for low-rank density matrices of the final state, such
a complex entanglement dynamics can be completely described by
factors representing single-sided entanglement dynamics of the
initial state. For two qubits, in particular, we show that the above
factorization can be achieved, if the rank of the final state
density matrix is two. For multiqubit systems, in contrast, the
factorization is possible for density matrices with rank no higher
then four. On analytical examples and by numerical simulations we
show that the factorization is independent on \cor{the initial
(pure) quantum state of the qubit system and the local channels}, as
far as the above rank conditions are fulfilled. Since it is
generally difficult to generate a low rank final state density
matrix out of an arbitrary initial multiqubit state, for three and
four qubit systems, we \cor{shall assume first} that the initial
state is a maximally entangled state, either GHZ or W state. The
symmetry of the initial states allows us generating final state
density matrices with all possible ranks \cor{for arbitrary local
channels. Later we shall relieve this latter assumption extending
our results to the case of arbitrary initially pure state of the
qubits.}

This work is organized as follows. In the next section we shall
briefly describe the entanglement measures of use and introduce the
quantum operation formalism \cite{Kraus:83,Nielsen:00} that allows
us to access the state dynamics of quantum systems under the action
of local noisy channels. In Sec.~\ref{sec:fac} we step-by-step
analyze the entanglement dynamics of two-, three- and four-qubit
systems affected by local many-sided channels and show examples when
such a complex entanglement dynamics can be factorized on terms
representing single-sided entanglement evolution. We conclude in
Sec.~\ref{sec:dis} with a summary of our results and a discussion of
their possible implications to theoretical and experimental
description of the entanglement dynamics.

\section{Concurrence and quantum state dynamics}

\subsection{\label{sec:ent meas} The entanglement measure}

It has been found difficult to quantify the entanglement of mixed
many-partite states, and no general solution is known
\cite{Horodecki:09} apart from Wootter's concurrence
\cite{Wootters:98}. The Wootter's concurrence allows us to compute
the entanglement of an arbitrary state of a two-qubit system, which
is given by the density matrix $\rho$, as $C_{W} = {\rm max} \{ 0,
\; \lambda^1 - \lambda^2 - \lambda^3 - \lambda^4\}$. Here
$\lambda^i$ are the square roots of the four eigenvalues of the
non-Hermitean matrix $\rho\: (\sigma_y \otimes \sigma_y ) \rho^\ast
(\sigma_y \otimes \sigma_y )$, if taken in decreasing order. It is
important to note that this matrix is obtained from the density
matrix $\rho$ by simultaneous inversion of the single-qubit
subsystems with the help of the only generator $\sigma_y$ of the
SO$(2)$ group.

Various extensions of Wootter's concurrence have been worked out
over the years \cite{Horodecki:09}. Ou \etal{} \cite{Ou:08}, in
particular, suggested a generalization of Wootter's concurrence for
bipartite states, if the dimensions of the associated Hilbert
subspaces are larger than two. For a $d_1\otimes d_2$-dimensional
quantum system, this concurrence can be written as
\begin{equation}
 \label{concurence-gen}
C = \sqrt{ \sum_{m=1}^{d_1(d_1-1)/2} \sum_{n=1}^{d_2(d_2-1)/2}
\left( C_{m n} \right)^2} \, ,
\end{equation}
where each term $C_{m n}$ is given by
\begin{equation}
 \label{concurence}
C_{m n} = {\rm max} \{ 0, \lambda_{m n}^1 - \lambda_{m n}^2 -
\lambda_{m n}^3 - \lambda_{m n}^4 \} \, .
\end{equation}
Here, the $\lambda_{m n}^k, \, k=1..4$ are the square roots of the
four nonvanishing eigenvalues of the matrix $\rho\, \tilde{\rho}_{m
n}$, if taken in decreasing order. These matrices $\rho\:
\tilde{\rho}_{m n}$ are formed by means of the density matrix $\rho$
and its complex conjugate $\rho^*$, and are further transformed by
the operators $ S_{m n} = L_m \otimes L_n$ as: $\tilde{\rho}_{m n} =
S_{m n} \rho^\ast S_{m n}$. In this notation, moreover, $L_m$ are
$d_1(d_1-1)/2$ generators of the group SO$(d_1)$, while the $ L_n$
are the $d_2(d_2-1)/2$ generators of the group SO$(d_2)$.

Although the bipartite concurrence (\ref{concurence-gen}) reduces to
Wootter's concurrence for the special case of two qubits, in general
it is an approximate entanglement measure, which provides limited
information about entanglement of the bipartite system \cite{Ou:08}.
While the dimensionality of the Hilbert space of a two-qubit system
is four, the inversion of an arbitrary state $\rho$ is unambiguously
defined by the single generator of the SO$(2)$ group. In higher
dimensions, however, there is no unique way to invert a given
quantum state \cite{Rungta:01}. Ambiguous choice of the state
inversion leads to the summation over all possible
$d_1(d_1-1)d_2(d_2-1)/4$ state inversions in
Eq.~(\ref{concurence-gen}) in all $2\otimes 2$-dimensional subspaces
of the original $d_1\otimes d_2$-dimensional Hilbert state space of
the bipartite system. The main consequence of such approximation for
state inversion is that there may be only four nonzero eigenvalues
of the matrix $\rho\, \tilde{\rho}_{m n}$, while the other $d_1 d_2
- 4$ eigenvalues of \cor{this matrix} always vanish.

In spite of the above limitations, concurrence
(\ref{concurence-gen}) has been shown to be quite powerful measure
of entanglement \cite{Li(d):09,Siomau:12a,Yu:08}. Using the
bipartite concurrence Li \etal \cite{Li(b):09} formulated an
analytical lower bound for multiqubit concurrence, which is given by
a squared sum of the bipartite concurrences computed for all
possible bi-partitioning of the multiqubit system. For three qubits,
in particular, the lower bound can be written in terms of the three
bipartite concurrences that correspond to possible cuts of two
qubits from the remaining one, i.e.
\begin{equation}
\label{low-bound-three}
 \tau_3 (\rho) = \sqrt{\frac{1}{3}\, \left( (C^{12|3})^2  + (C^{13|2})^2 + (C^{23|1})^2 \right)} \, .
\end{equation}
This lower bound, moreover, has been used to describe the
entanglement dynamics of three-qubit states under the action of
certain multi-sided noisy channels \cite{Siomau:10a,Siomau:11}. On
particular analytical examples and by numerical simulations, it has
been shown that for three-qubit density matrices with rank no higher
then four, the lower bound (\ref{low-bound-three}) provides adequate
description of the entanglement evolution irrespective from
system-channel coupling rate and for all times of interaction. For
density matrices with higher ranks, however, the lower bound
vanishes after a finite time, while the quantum states it is applied
to are not separable, i.e. possess certain amount of entanglement.
This behavior of the lower bound (\ref{low-bound-three}) is not the
consequence of the entanglement sudden death \cite{Yu:06}, but is
induced by the approximate character of the bipartite concurrence
(\ref{concurence-gen}) as an entanglement measure.

\subsection{Quantum Operation Formalism}

Quantum operation formalism is a very general and prominent tool to
describe how a quantum system has been influenced by its
environment. According to this formalism the final state of the
quantum system, that is coupled to some environmental channel, can
be obtained from its initial state with the help of (Kraus)
operators
\begin{equation}
 \label{sum-represent}
 \rho_{\rm fin} = \sum_i K_i \, \rho_{\rm ini} \, K_i^\dag \, ,
\end{equation}
and the condition $\sum_i K_i^\dag \, K_i = I$ is fulfilled. Note
that we consider only such system-environment interactions that can
be associated with completely positive trace-preserving maps
\cite{Nielsen:00}.

If the quantum system of interest consists of just \cor{a single
qubit,} which is subjected to some environmental channel $A$, then
an arbitrary quantum operation associated with the channel's action
can be expressed with the help of at most four operators
\cite{Nielsen:00}. Let us define the four operators through the
Pauli matrices as
\begin{eqnarray}
  \label{single}
 K_1(a_1) &=& \frac{a_1}{\sqrt{2}} \left( \begin{array}{cc} 1 & 0 \\ 0 & 1
\end{array} \right) , \;
 K_2(a_2) = \frac{a_2}{\sqrt{2}} \left( \begin{array}{cc} 0 & 1 \\1 &
0 \end{array} \right) \, ,
 \\[0.1cm]
 K_3(a_3) &=& \frac{a_3}{\sqrt{2}} \left( \begin{array}{cc} 0 & -i \\ i & 0
\end{array} \right) , \:
 K_4(a_4) = \frac{a_4}{\sqrt{2}} \left( \begin{array}{cc} 1 & 0 \\0 & -1
\end{array} \right) \, , \nonumber
\end{eqnarray}
where $a_i$ are real parameters and the condition $\sum_{i=1 }^4
a_i^2 = 1$ holds. Following standard notations \cite{Nielsen:00},
the channel with parameters $a_1\neq 0, \, a_2\neq 0$ and
$a_3=a_4=0$ is called bit flip (BF); if $a_1\neq 0, \, a_4\neq 0$
and $a_2=a_3=0$ the channel is called phase flip (PF), while the
channel with $a_1\neq 0, \, a_3\neq 0$ and $a_2=a_4=0$ is bit-phase
flip (BPF). These notations and definitions shall be used along the
paper.

\section{\label{sec:fac} Factorization of concurrence}

As we mentioned in the Introduction, our ability to simplify the
description of the entanglement evolution strongly depends on the
symmetries of the qubits-environment system. Therefore, \cor{at the
beginning} we shall assume that the quantum system is initially
prepared in a maximally entangled state, i.e. in a Bell state for
two qubits and either in GHZ or W state for three and four qubits.
At the same time we do not impose any limitations on the
single-qubit channels, apart from their representation by operators
(\ref{single}). Moreover, if in the text, for example, two
single-qubit channels $A$ and $B$ are both BF channels, it is
assumed that in general these channels are not equivalent, i.e.
$a_1\neq b_1$ and $a_2\neq b_2$.

\subsection{Two qubits}

Suppose, each qubit of the two-qubit system, initially prepared into
a Bell state $\ket{\varphi} = 1/\sqrt{2} \left( \ket{00}+\ket{11}
\right)$, is affected by the BF channels $A$ and $B$. Using the
definitions for the BF channel (\ref{single}) and Wootter's
concurrence we obtain that
\begin{eqnarray}
 \label{2-q fac}
C_W ([A \otimes B] \ket{\varphi}\bra{\varphi}) &=&
   \nonumber
   \\[0.1cm]
   & & \hspace*{-2.5cm} C_W ([1 \otimes A]
\ket{\varphi}\bra{\varphi}) \; C_W ([1 \otimes B]
\ket{\varphi}\bra{\varphi}) \, ,
\end{eqnarray}
i.e. the entanglement dynamics of the two-qubit system affected by
the two-sided channel $[A \otimes B]$ is completely defined by the
action of single-sided channels $[1 \otimes A]$ and $[1 \otimes B]$
on the initial state $\ket{\varphi}$. It is easy to verify that the
factorization (\ref{2-q fac}) holds true also if both channels $A$
and $B$ are PF or BPF.

In all the cases above the final state density matrix $[A \otimes B]
\ket{\varphi}\bra{\varphi}$ has rank two. We have numerically
generated channels $A$ and $B$ and confirmed that for any final
density matrix with rank two, the factorization (\ref{2-q fac})
remains valid irrespective of the channels $A$ and $B$.
\cor{Moreover, the factorization (\ref{2-q fac}) is possible even
when the initial state is not the maximally entangled state, but an
arbitrary pure state. For example, one may verify that if
$\ket{\varphi} = 1/\sqrt{2} \left( \alpha \ket{00} +
\sqrt{1-\alpha^2} \ket{11} \right)$ and the channels $A$ and $B$ are
both BF, the final state density matrix is of rank two and
Eq.~(\ref{2-q fac}) is still a valid decomposition. For nonmaximally
entangled initial states $\ket{\varphi}$, however, an additional
factor $C_W(\varphi)$ appears in the right hand side of
Eq.~(\ref{2-q fac}). This factor describes only the initial
entanglement of these states and does not cause any impact on the
description of the entanglement dynamics.}

\subsection{Three qubits}

In contrast to two qubits, there are two maximally entangled states
for three-qubit systems \cite{Duer:00}, which are typically written
\cor{in the computational basis} as
\begin{eqnarray}
  \label{GHZ}
&& \ket{\rm GHZ_3} = \frac{1}{\sqrt{2}} \left( \ket{000} + \ket{111}
\right) \, , \\[0.1cm]
  \label{W}
&& \ket{\rm W_3} = \frac{1}{\sqrt{3}} \left( \ket{001} + \ket{010} +
\ket{100} \right) \, .
\end{eqnarray}

The entanglement dynamics of these states in local channels can be
described, for example, by the bipartite concurrence $C^{12|3}$ that
correspond to the separation of two qubits of the three-qubit system
from the remaining one. Let us consider the three-qubit system
initially prepared in the GHZ state (\ref{GHZ}). If each of the
qubits is simultaneously affected by the PF channels, the final
state $[A \otimes B \otimes C] \ket{\rm GHZ_3}\bra{\rm GHZ_3}$ is
given by rank-2 density matrix. In this case, the bipartite
concurrence $C^{12|3}$ can be factorized in a way similar to
Eq.~(\ref{2-q fac}), i.e.
\begin{eqnarray}
 \label{3-q fac}
C^{12|3} ([A \otimes B \otimes C] \ket{\psi}\bra{\psi}) &=&
\nonumber \\[0.1cm]
&& \hspace*{-2cm} C^{12|3}(\mathcal{A}) \;\; C^{12|3}(\mathcal{B})
\;\; C^{12|3}(\mathcal{C}) \, ,
\end{eqnarray}
where
\begin{eqnarray}
  \label{ABC}
\mathcal{A} &\equiv& \left[1\otimes 1\otimes
A\right]\ket{\psi}\bra{\psi} \, , \nonumber \\[0.1cm]
\mathcal{B} &\equiv& \left[1\otimes 1\otimes
B\right]\ket{\psi}\bra{\psi} \, , \nonumber \\[0.1cm]
\mathcal{C} &\equiv& \left[1\otimes 1\otimes
C\right]\ket{\psi}\bra{\psi} \, ,
\end{eqnarray}
where $\ket{\psi}$ is the initial state.

If, in contrast, the three qubits are initially prepared into the
GHZ state (\ref{GHZ}) and are all simultaneously affected by BF
channels, or two of the qubits undergo the action of PF channels,
while the remaining one is the subject of BF or BPF, the final state
density matrix has rank four and the bipartite concurrence
$C^{12|3}$ can be factorized as
\begin{eqnarray}
 \label{3-q fac adv}
C^{12|3} ([A \otimes B \otimes C] \ket{\psi}\bra{\psi}) &=&
\nonumber \\[0.1cm]
&& \hspace*{-4cm} C^{12|3}(\mathcal{A}) \; C^{12|3}(\mathcal{C}) \;
+ \; C^{12|3}(\mathcal{B}) \; C^{12|3}(\mathcal{C}) \, .
\end{eqnarray}
This factorization (\ref{3-q fac adv}) remains valid also for the
rank-3 density matrix $[A \otimes B \otimes C] \ket{\rm W_3}\bra{\rm
W_3}$, where three PF channels affect the initial W state (\ref{W}).
By numerical generation of the channels $A, B$ and $C$, we checked
that the factorization (\ref{3-q fac adv}) is true for arbitrary
channels $A, B$ and $C$ affecting one of the initial states
(\ref{GHZ}) or (\ref{W}), if the rank of the final state density
matrix is three or four. \cor{We also obtained a few numerical
evidences that the factorizations (\ref{3-q fac}) and (\ref{3-q fac
adv}) are true even for nonmaximally entangled pure states of
three-qubit systems. As in the case of two qubits, which we
discussed in the previous section, an additional factor appears in
the right hand side of Eqs.~(\ref{3-q fac}) and (\ref{3-q fac adv}),
if the initial state is not a maximally entangled state. This factor
gives the initial entanglement of the three-qubit state.}

Eqs.~(\ref{3-q fac}) and (\ref{3-q fac adv}) have very different
structure of the right hand side. This can be understood from the
definition of the bipartite concurrence (\ref{concurence-gen}). As
we mentioned in Sec.~\ref{sec:ent meas}, because of lack of the
unique state inversion in higher dimensions, the bipartite
concurrence takes into account all possible state inversions in all
$2\otimes 2$-dimensional subspaces of the original Hilbert state
space. The bipartite concurrence $C^{12|3}$ separates the
three-qubit system on two subsystems, so that two of three qubits
are unified in one subsystem. If the GHZ state (\ref{GHZ}) is
affected by three PF channels $A, B$ and $C$, both the two-qubit
subsystem affected by channels $A$ and $B$, and the single-qubit
subsystem affected by channel $C$ lie in two-dimensional subspaces
of the Hilbert space of three-qubits. In this special case the state
inversion is unique, and there is only single term in the right hand
side of Eq.~(\ref{3-q fac}). If, in contrast, the GHZ state
(\ref{GHZ}) is affected by three channels $A, B$ and $C$, so that
the final state density matrix has rank four, the two-qubit
subsystem subjected to the action of the channels $A$ and $B$
belongs to a four-dimensional subspace, while the single-qubit
subsystem affected by channel $C$ is the subject of a
two-dimensional subspaces. This leads to two terms in the
factorization (\ref{3-q fac adv}) caused by unambiguous choice of
the two-dimensional subspace from the four-dimensional subspace of
the two-qubit subsystem.

\subsection{Four qubits}

Entanglement dynamics of three-qubit systems can be quantified only
with the bipartite concurrence $C^{12|3}$ where two qubits are
separated from the remaining one. For four-qubit system, in
contrast, we may consider two bipartite concurrences $C^{123|4}$ and
$C^{12|34}$ and have a better insight of the multiqubit entanglement
dynamics in many-sided channels and possible ways to simplify its
description through the factorization. As in the previous section we
shall focus on the two maximally entangled states, i.e.
\begin{eqnarray}
  \label{GHZ-4}
&& \ket{\rm GHZ_4} = \frac{1}{\sqrt{2}} \left( \ket{0000} +
\ket{1111}
\right) \, , \\[0.1cm]
  \label{W-4}
&& \hspace*{-1cm} \ket{\rm W_4} = \frac{1}{2} \left( \ket{0001} +
\ket{0010} + \ket{0100} + \ket{1000} \right) \, .
\end{eqnarray}

Let us assume that the four-qubit GHZ state (\ref{GHZ-4}) is
affected by four PF channels. The final state density matrix $[A
\otimes B \otimes C \otimes D] \ket{\rm GHZ_4}\bra{\rm GHZ_4}$ has
rank two. In this special case both bipartite concurrences
$C^{123|4}$ and $C^{12|34}$ factorize as
\begin{eqnarray}
 \label{4-q-1 fac}
C^{123|4} ([A \otimes B \otimes C \otimes D] \ket{\psi}\bra{\psi}) =
\nonumber \\[0.1cm]
&& \hspace*{-5cm} C^{123|4}(\mathcal{A}) \; C^{123|4}(\mathcal{B})
\; C^{123|4}(\mathcal{C})\; C^{123|4}(\mathcal{D}) \, ,
\nonumber \\[0.2cm]
\label{4-q-2 fac} C^{12|34} ([A \otimes B \otimes C \otimes D]
\ket{\psi}\bra{\psi}) =
\nonumber \\[0.1cm]
&& \hspace*{-5cm} C^{12|34}(\mathcal{A}) \; C^{12|34}(\mathcal{B})
\; C^{12|34}(\mathcal{C})\; C^{12|34}(\mathcal{D}) \, ,
\end{eqnarray}
where $\mathcal{D}$ is defined by analogy with Eqs.~(\ref{ABC}) as
$\mathcal{D} \equiv \left[1\otimes 1\otimes 1\otimes
D\right]\ket{\psi}\bra{\psi}$. This factorization mimics the
structure of Eqs.~(\ref{2-q fac}) and (\ref{3-q fac}), which were
obtained for two- and three-qubit systems, i.e. there is only single
term in the right hand side of this equation.

If, however, the final state density matrix has rank four, the
concurrences $C^{123|4}$ and $C^{12|34}$ factorize differently. The
rank four density matrix can be obtained, for example, by the action
of three PF channels and single BF channel on the initial GHZ state
(\ref{GHZ-4}) or by action of four PF channels on W state
(\ref{W-4}). In these cases we have
\begin{eqnarray}
 \label{4-q-1 fac 2vs2}
C^{123|4} ([A \otimes B \otimes C \otimes D] \ket{\psi}\bra{\psi}) =
\nonumber \\[0.1cm]
&& \hspace*{-5cm} C^{123|4}(\mathcal{A}) \; C^{123|4}(\mathcal{D})
\; + \; C^{123|4}(\mathcal{B})\; C^{123|4}(\mathcal{D})
\nonumber \\[0.1cm]
&& \hspace*{-4cm} \; + \; C^{123|4}(\mathcal{C})\;
C^{123|4}(\mathcal{D})\, ,
\\[0.4cm]
\label{4-q-2 fac 2vs2} C^{12|34} ([A \otimes B \otimes C \otimes D]
\ket{\psi}\bra{\psi}) =
\nonumber \\[0.1cm]
&& \hspace*{-5.5cm} C^{12|34}(\mathcal{A}) \; C^{12|34}(\mathcal{C})
\; + \; C^{12|34}(\mathcal{A})\; C^{12|34}(\mathcal{D})
\nonumber \\[0.1cm]
&& \hspace*{-5.8cm} \; + \; C^{12|34}(\mathcal{B})\;
C^{12|34}(\mathcal{C}) \; + \; C^{12|34}(\mathcal{B})\;
C^{12|34}(\mathcal{D}) \, .
\end{eqnarray}
The structure of these equations is similar to Eq.~(\ref{3-q fac
adv}) for three qubits and can be explained using the interpretation
given in the previous section: each factor in the right had side of
these equations represent a particular selection of a $2\otimes
2$-dimensional subspace of the original four-qubit state space.

Because a general four-qubit mixed state density matrix has rank 16,
it is difficult to generate low-rank density matrices even from the
maximally entangled states. \cor{Therefore, very few numerical
results have been obtained to confirm the factorizations (\ref{4-q-1
fac 2vs2})-(\ref{4-q-2 fac 2vs2}) for an arbitrary local channels
and only for initial maximally entangled states (\ref{GHZ-4}) and
(\ref{W-4}).}

\section{\label{sec:dis} Results and Discussion}

We have shown that the complex entanglement evolution of multiqubit
systems in many-sided noisy channels can be factorized on terms
representing the entanglement evolution in single-sided channels. It
has been argued and confirmed on analytical examples and by
numerical simulations that this factorization is independent on
\cor{the local noisy channels and initial pure states of the qubit
systems, and is solely defined by the rank of the final state
density matrices. If the rank of the final state density matrix is
two [cf.~Eqs.~(\ref{2-q fac}), (\ref{3-q fac}) and (\ref{4-q-1
fac})], the complex multi-sided entanglement dynamics is given by a
product of factors} describing the single-sided entanglement
evolution. For rank-3 and rank-4 density matrices
[cf.~Eqs.~(\ref{3-q fac adv}), (\ref{4-q-1 fac 2vs2}) and
(\ref{4-q-2 fac 2vs2})], in contrast, the entanglement dynamics is
represented by a sum of terms giving different combinations of
products representing the single-sided entanglement evolution.

In all the factorization equations mentioned above, the ranks of the
density matrices in the right hand side are no higher then the ranks
of the density matrices in the left hand side. This means that these
equations provide low-rank decompositions of the mixed state density
matrix with respect to (Wootter's or bipartite) concurrence. These
decompositions can be, in principle, used to approximate the complex
entanglement dynamics, when the rank of the final state density
matrix is higher then four. If, for example, the initial three-qubit
GHZ state (\ref{GHZ}) is affected by three BPF channels $A, B$ and
$C$ (for simplicity we assume that these channels are identical,
i.e. $a_1=b_1=c_1=\sqrt{1-p}$ and $a_3=b_3=c_3=\sqrt{p}$), the final
state density matrix $[A \otimes B \otimes C] \ket{\rm
GHZ_3}\bra{\rm GHZ_3}$ has rank 8. The entanglement dynamics of this
three-qubit state can be directly described with the lower bound
(\ref{low-bound-three}). If, in contrast, we assume that the right
hand side of Eq.~(\ref{3-q fac}) is a valid decomposition for each
bipartite concurrence $C^{ab|c}$ (where $a,b,c=1..3$ and $a\neq
b\neq c\neq a$) in the definition of the lower bound
(\ref{low-bound-three}), the multi-sided entanglement dynamics is
simply given by $\tau_3 \left( [A \otimes B \otimes C] \ket{\rm
GHZ_3}\bra{\rm GHZ_3} \right) = (1-2p)^3$. If, in contrast, we
assume that the right hand side of Eq.~(\ref{3-q fac adv}) is a
valid decomposition for the bipartite concurrence, then $\tau_3
\left( [A \otimes B \otimes C] \ket{\rm GHZ_3}\bra{\rm GHZ_3}
\right) = (1-2p)^2$.

\begin{figure}
\begin{center}
\includegraphics[scale=0.7]{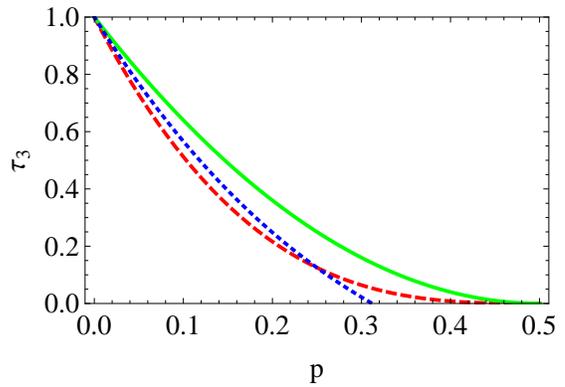}
\caption{(Color online) Entanglement evolution of the three-qubit
GHZ state affected by three BPF channels: direct application of the
lower bound (\ref{low-bound-three}) dotted blue; the approximation
with help of the right hand side of Eq.~(\ref{3-q fac}) dashed red
and the approximation with the right hand side of Eq.~(\ref{3-q fac
adv}) solid green.}
 \label{fig}
\end{center}
\end{figure}

Fig.~\ref{fig} shows the comparison between the direct application
of the lower bound to the final state density matrix $[A \otimes B
\otimes C] \ket{\rm GHZ_3}\bra{\rm GHZ_3}$ and the behavior of the
lower bound taking into account decompositions (\ref{3-q fac}) and
(\ref{3-q fac adv}). The approximations of the lower bound with help
of the factorizations (\ref{3-q fac}) and (\ref{3-q fac adv})
provide adequate description of the entanglement dynamics of the
three-qubit state. Moreover, the lower bound vanish for $p\approx
0.31$ due to the drawback of its construction, which we discussed in
Sec.~\ref{sec:ent meas}, and because of its application to the
rank-8 density matrix $[A \otimes B \otimes C] \ket{\rm
GHZ_3}\bra{\rm GHZ_3}$. At the same time the approximations
exploiting Eqs.~(\ref{3-q fac}) and (\ref{3-q fac adv}) nullify only
for $p=0.5$. Using a separability criteria \cite{Siomau:10a} one may
check that the state $[A \otimes B \otimes C] \ket{\rm
GHZ_3}\bra{\rm GHZ_3}$, where the channels are the BPF, become
separable only for $p=0.5$. Thus, the factorization approximations
provide even better description of the \cor{long-time} entanglement
dynamics then the original lower bound (\ref{low-bound-three}) in
applications to high-rank mixed state density matrices.

The main difficulty to construct the approximations using
Eqs.~(\ref{3-q fac}) and (\ref{3-q fac adv}) is to choose which of
these equations should be used for the approximation, since they
provide different description of the entanglement dynamics. In
principle, it should be possible to construct an entanglement
measure $E$ that approximates a complex multi-sided entanglement
dynamics $E \left( [A \otimes ... \otimes Z] \ket{\psi}\bra{\psi}
\right)$ by a function of terms $E(\mathcal{A}),...,E(\mathcal{Z})$
describing the single-sided entanglement dynamics, where
$\mathcal{A},...,\mathcal{Z}$ are defined by analogy with
Eq.~(\ref{ABC}). However, the construction of such an entanglement
measure requires further study and, therefore, shall be discussed
elsewhere.

Finally, the derived factorization equations as well as the
approximations to the entanglement dynamics based on these equations
can substantially simplify experimental detection of the
entanglement of a system undergoing complex multi-sided entanglement
dynamics as well as experimental testing of realistic communication
channels. It is known that the complexity of an entanglement witness
(i.e. the number of local measurements) \cite{Guehne:09} strongly
depends on the rank of the density matrix representing the quantum
state of interest. The factorization of the complex multi-sided
entanglement dynamics on terms providing single-sided entanglement
evolution allows reducing the complexity of the corresponding
entanglement witnesses and thus simplify the procedure of
entanglement detection and experimental quantitative description.

\begin{acknowledgments}
This work is supported by KACST under research grant no.~34-37.
\end{acknowledgments}

\end{document}